\documentclass[aps,prl,superscriptaddress,longbibliography,twocolumn]{revtex4-1}
\usepackage{graphicx}
\usepackage{color}	
\usepackage{natbib}
	\usepackage{url}

\begin{document}

%\preprint{}

\title{Direct visualization of phase separation between superconducting and nematic domains in Co-doped CaFe$_2$As$_2$ close to a first order phase transition}

\author{Ant\'on Fente}
\affiliation{Laboratorio de Bajas Temperaturas, Departamento de F\'isica de la Materia
Condensada, Instituto Nicol\'as Cabrera and
Condensed Matter Physics Center (IFIMAC), Universidad Aut\'onoma de Madrid, E-28049 Madrid,
Spain}

\author{Alexandre Correa-Orellana}
\affiliation{Instituto de Ciencia de Materiales de Madrid, Consejo Superior de Investigaciones Cient\'{\i}ficas (ICMM-CSIC), Sor Juana In\'es de la Cruz 3, 28049 Madrid, Spain}

\author{Anna E. B\"ohmer}
\affiliation{Department of Physics and Astronomy and Ames Laboratory, U.S. DOE, Iowa State University, Ames, Iowa 50011, USA}

\author{Andreas Kreyssig}
\affiliation{Department of Physics and Astronomy and Ames Laboratory, U.S. DOE, Iowa State University, Ames, Iowa 50011, USA}

\author{S. Ran}
\affiliation{Department of Physics and Astronomy and Ames Laboratory, U.S. DOE, Iowa State University, Ames, Iowa 50011, USA}

\author{Sergey~L. Bud'ko}
\affiliation{Department of Physics and Astronomy and Ames Laboratory, U.S. DOE, Iowa State University, Ames, Iowa 50011, USA}

\author{Paul~C. Canfield}
\affiliation{Department of Physics and Astronomy and Ames Laboratory, U.S. DOE, Iowa State University, Ames, Iowa 50011, USA}

\author{Federico Mompean}
\affiliation{Instituto de Ciencia de Materiales de Madrid, Consejo Superior de Investigaciones Cient\'{\i}ficas (ICMM-CSIC), Sor Juana In\'es de la Cruz 3, 28049 Madrid, Spain}
\affiliation{Unidad Asociada de Bajas Temperaturas y Altos Campos Magn\'eticos, UAM, CSIC, Cantoblanco, E-28049 Madrid, Spain}

\author{Mar Garc{\'i}a-Hern{\'a}ndez}
\affiliation{Instituto de Ciencia de Materiales de Madrid, Consejo Superior de Investigaciones Cient\'{\i}ficas (ICMM-CSIC), Sor Juana In\'es de la Cruz 3, 28049 Madrid, Spain}
\affiliation{Unidad Asociada de Bajas Temperaturas y Altos Campos Magn\'eticos, UAM, CSIC, Cantoblanco, E-28049 Madrid, Spain}

\author{Carmen Munuera}
\affiliation{Instituto de Ciencia de Materiales de Madrid, Consejo Superior de Investigaciones Cient\'{\i}ficas (ICMM-CSIC), Sor Juana In\'es de la Cruz 3, 28049 Madrid, Spain}
\affiliation{Unidad Asociada de Bajas Temperaturas y Altos Campos Magn\'eticos, UAM, CSIC, Cantoblanco, E-28049 Madrid, Spain}

\author{Isabel Guillam\'on}
\affiliation{Laboratorio de Bajas Temperaturas, Departamento de F\'isica de la Materia
Condensada, Instituto Nicol\'as Cabrera and
Condensed Matter Physics Center (IFIMAC), Universidad Aut\'onoma de Madrid, E-28049 Madrid,
Spain}
\affiliation{Unidad Asociada de Bajas Temperaturas y Altos Campos Magn\'eticos, UAM, CSIC, Cantoblanco, E-28049 Madrid, Spain}

\author{Hermann Suderow}
\affiliation{Laboratorio de Bajas Temperaturas, Departamento de F\'isica de la Materia
Condensada, Instituto Nicol\'as Cabrera and
Condensed Matter Physics Center (IFIMAC), Universidad Aut\'onoma de Madrid, E-28049 Madrid,
Spain}
\affiliation{Unidad Asociada de Bajas Temperaturas y Altos Campos Magn\'eticos, UAM, CSIC, Cantoblanco, E-28049 Madrid, Spain}


\date{\today}

\begin{abstract}
We show that biaxial strain induces alternating tetragonal superconducting and orthorhombic nematic domains in Co substituted CaFe$_2$As$_2$. We use Atomic Force, Magnetic Force and Scanning Tunneling Microscopy (AFM, MFM and STM) to identify the domains and characterize their properties, finding in particular that tetragonal superconducting domains are very elongated, more than several tens of $\mu$m long and about 30 nm wide, have the same $T_c$ than unstrained samples and hold vortices in a magnetic field. Thus, biaxial strain produces a phase separated state, where each phase is equivalent to what is found at either side of the first order phase transition between antiferromagnetic orthorhombic and superconducting tetragonal phases found in unstrained samples when changing Co concentration. Having such alternating superconducting domains separated by normal conducting domains with sizes of order of the coherence length opens opportunities to build Josephson junction networks or vortex pinning arrays and suggests that first order quantum phase transitions lead to nanometric size phase separation under the influence of strain.
\end{abstract}

%\keywords{}

\maketitle


%\section{Annealing and phase diagram}
%\subsection{}
%\subsubsection{}

	\begin{figure}
		\begin{center}
			\includegraphics[clip=true,width=0.45\textwidth,keepaspectratio]{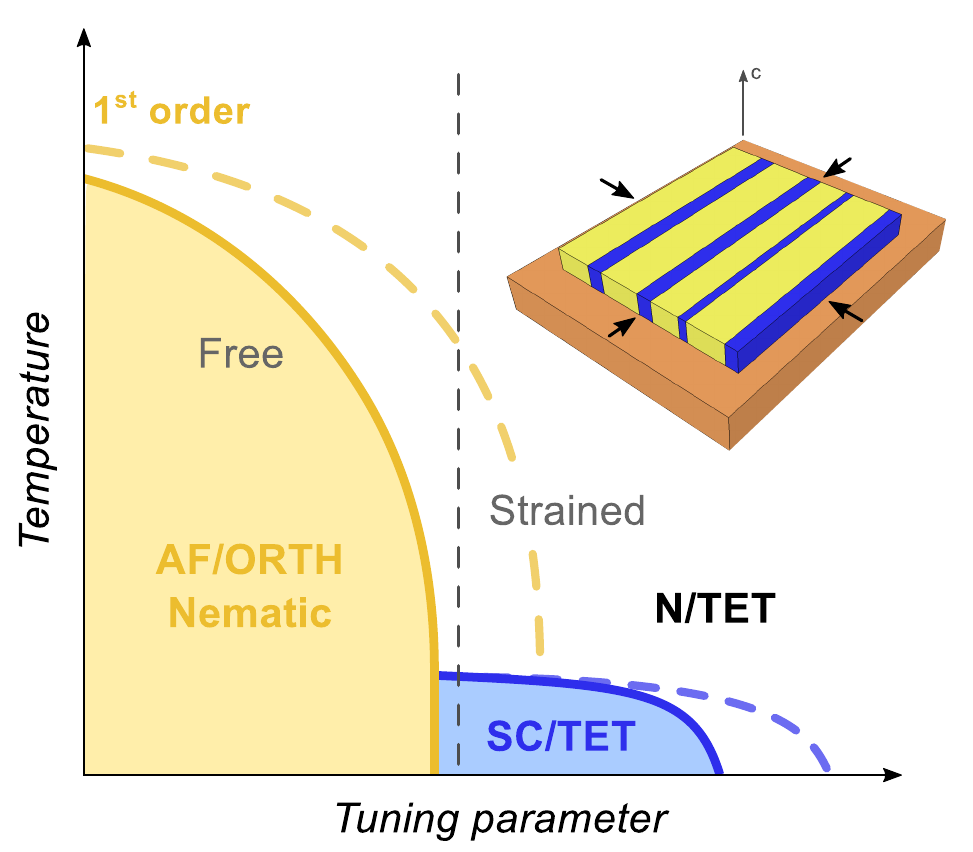}
		\end{center}
		\vskip -.4 cm
		\caption{Generic phase diagram, with a first order tetragonal/orthorhombic phase transition and superconductivity (blue lines) appearing when the AF/ORTH (yellow lines) order disappears. Lines are for free standing samples and dashed lines for strained samples. In the upper right panel we schematically show the configuration we find. Biaxial strain (dark arrows) is applied by using the different thermal expansion of the copper support (orange rectangle in the figure) and the sample (yellow/blue rectangle in the figure). Biaxial strain results, for $x=3.2$\% studied here, in alternating tetragonal superconducting (blue) and antiferromagnetic orthorhombic (yellow) domains. The superconducting domains are long stripes a few tens of $nm$ wide and hold vortex lattices.}
		\label{DiagramaFases}
	\end{figure}

\section{Introduction}

The Ca(Fe$_{1-x}$Co$_x$)$_2$As$_2$ system stands out as one of the most pressure and strain sensitive inorganic compounds. Using hydrostatic pressure, Gati et al.\,were able to show that the transition temperature $T_{s-m}$ of the first order coupled structural and magnetic transition between the high temperature tetragonal paramagnetic state to the low temperature orthorhombic antiferromagnetic state decreases with pressure as d$T_{s-m}/$d$P\approx - 1100$ K/GPa and that the superconducting transition temperature $T_c$ has d$T_{c}/$d$P\approx - 60$K/GPa\cite{Gati2012}. Both numbers are huge, stressing the strong sensitivity of superconducting, magnetic and structural properties to minute modifications of lattice parameters\cite{Ran2011,Ran2012}.

Ca(Fe$_{1-x}$Co$_x$)$_2$As$_2$ is also remarkably strain sensitive, manifesting clear shifts in the transition temperatures when subject to biaxial strain, as schematically represented in Fig.\,1\cite{Boehmer2017}. It has been conjecturized that for a range of Co concentrations $x$, the sample accomodates strain by microscopically breaking up into electronically different domains with, at low temperatures, separated regions of non-superconducting orthorhombic antiferromagnetic phase alternating with regions of superconducting, tetragonal and paramagnetic phase. B\"ohmer et al\,\cite{Boehmer2017} showed that biaxial strain is applied to the Ca(Fe$_{1-x}$Co$_x$)$_2$As$_2$ when it is firmly bonded to a substrate. They found that initially paramagnetic samples showed a structural transition and that initially non-superconducting samples became superconducting under strain. From X-ray scattering, they were also able to infer that biaxial strain must induce some sort of microscopic phase separation in Ca(Fe$_{1-x}$Co$_x$)$_2$As$_2$, making the transition gradual, contrary to results obtained when applying pressure or stress. However, no spatially resolved information about the different phases was obtained. Here we study a sample of Ca(Fe$_{1-x}$Co$_x$)$_2$As$_2$ for $x$ having the highest $T_c$ within the region where phase coexistence was found in Ref.\,\cite{Boehmer2017} using AFM, MFM and STM. We study the nature of each phase separately and show that locally the properties are very similar to the phases found in unstrained single-phase samples at different $x$---the normal orthorhombic antiferromagnetic phase shows the nematic electronic dispersion characteristic of the orthorhombic phase and the tetragonal superconducting phase shows the anticipated superconducting gap value and $T_c$\cite{Chuang2010,Ran2011}.

	\begin{figure}
		\begin{center}
			\includegraphics[clip=true,width=0.45\textwidth,keepaspectratio]{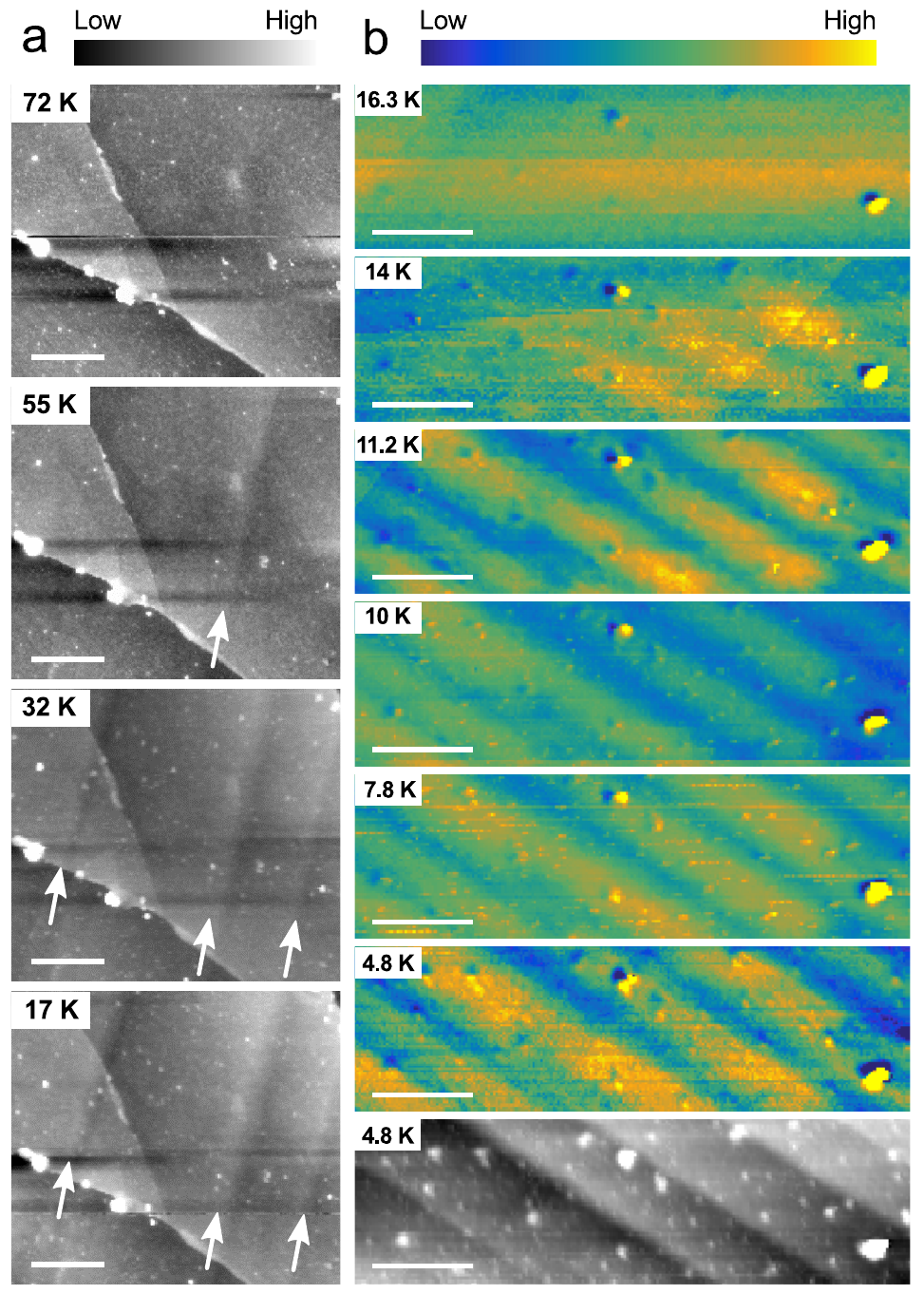}
		\end{center}
		\vskip -.4 cm
		\caption{Correlated structural and diamagnetic features. AFM (left column) and MFM (right column) data in Ca(Fe$_{1-x}$Co$_{x}$)$_2$As$_2$, $x = 0.032$ under biaxial strain. Each column corresponds to a different field of view. In the left column (AFM) we focus on the temperature dependence up to the structural transition. We observe stripes (marked by white arrows) that disappear when heating above $T_{s-m}\approx 70$\,K. Scale bars are of $2\,\mu m$. The AFM features do not evolve for temperatures below 17 K. In the right column we focus on the temperature dependence of the superconducting signal, measured by MFM at a magnetic field of 230 Oe. We observe stripes of varying magnetization. Diamagnetic signal corresponds to the blue part of the color scale. White scale bars are of $1.5\,\mu m$. In the bottom panel we show the AFM image corresponding to the MFM images in the panels above.}
		\label{Transiciones}
	\end{figure}	

\section{Experimental}

We study a single crystal doped with $3.2\,\%$ of Co and annealed at $350^{\circ}$C, as described in Ref.\cite{Ran2012}. The sample is about 0.1 mm thick and is firmly glued to a copper substrate. The thermal expansion leads to lateral length changes of about +0.5\% in the sample and -0.3\% in the copper substrate, leading to a differential thermal expansion of about 0.8\%\cite{Boehmer2017}. Our sample is located at the position indicated by the grey vertical dashed line in the schematic phase diagram of Fig.\ref{DiagramaFases}. To make STM experiments we use a home-made set-up similar to the one described in Ref.\cite{Suderow2011}. We cleave the samples \textit{in situ} at $4.2\,$K by gluing a piece of metal on top of the sample and moving the sample holder below a beam, in such a way as to push the piece of metal. Cleaving occurs in the $c$-axis and exposes large atomically flat surfaces. These are often atomically flat over the whole size of a single field of view of the STM set-up ($1.8\,\mu$m $\times 1.8\,\mu$m). Thanks to the positioning system described in Ref.\cite{Suderow2011}, we study many different fields of view. We perform the STM measurements using standard parameters (current of a few nA and bias voltages of a few tens of mV) and record topography in constant current mode as well as tunneling conductance maps to follow both structural features and the spatial dependence of the superconducting gap. After the STM measurements, we took the sample to a combined AFM/MFM system described in Ref.\cite{Galvis2015}. AFM and MFM measurements are made simultaneously using a cantilever with a pyramidal tip covered by a CrCo alloy. The AFM signal is taken as a function of the position by measuring the resonant properties of the cantilever. The cantilever is then retracted to a large distance (120 nm) to eliminate the signal resulting from surface interaction forces. To take the MFM image we maintain the frequency of the cantilever fixed using a feedback loop and record the phase shift induced by the magnetic force acting on the cantilever vs. $x$ and $y$\cite{Galvis2015}.

\section{Results}

At high temperatures in the tetragonal phase, AFM topography shows flat surfaces with terraces separated by steps a few unit cells high that are oriented randomly (see left column of Fig.\ref{Transiciones}). Below the partial transition into the orthorhombic phase (yellow dashed line in the phase diagram of Fig.\ref{DiagramaFases}) we observe stripes appearing on the surface (white arrows in left column of Fig.\ref{Transiciones}). They remain at the same position when cooling, and the corresponding contrast is enhanced.

At temperatures above $T_c = 16\,$K (blue dashed line in Fig.\ref{DiagramaFases}), MFM gives flat images without a clear spatial variation. In the superconducting phase, however, MFM maps show alternating diamagnetic and paramagnetic elongated regions (right column of Fig.\ref{Transiciones}) which remain at the same positions when cooling. The positions of the alternating diamagnetic and paramagnetic regions are correlated with the stripes observed in AFM. Flat terraces in AFM mostly provide paramagnetic signal. Diamagnetism in MFM is mostly restricted to places where we observe a linear feature in AFM.

It is useful to remark that the observed temperatures for the structural and superconducting transitions coincide with the ones found in Ref.\cite{Boehmer2017} for the same composition ($x$). This ascertains that we apply biaxial strain in very much the same way, by enforcing the deformation of the sample through firm bonding to the substrate.
	
\begin{figure}[htb]
		\begin{center}
			\includegraphics[clip=true,width=0.37\textwidth,keepaspectratio]{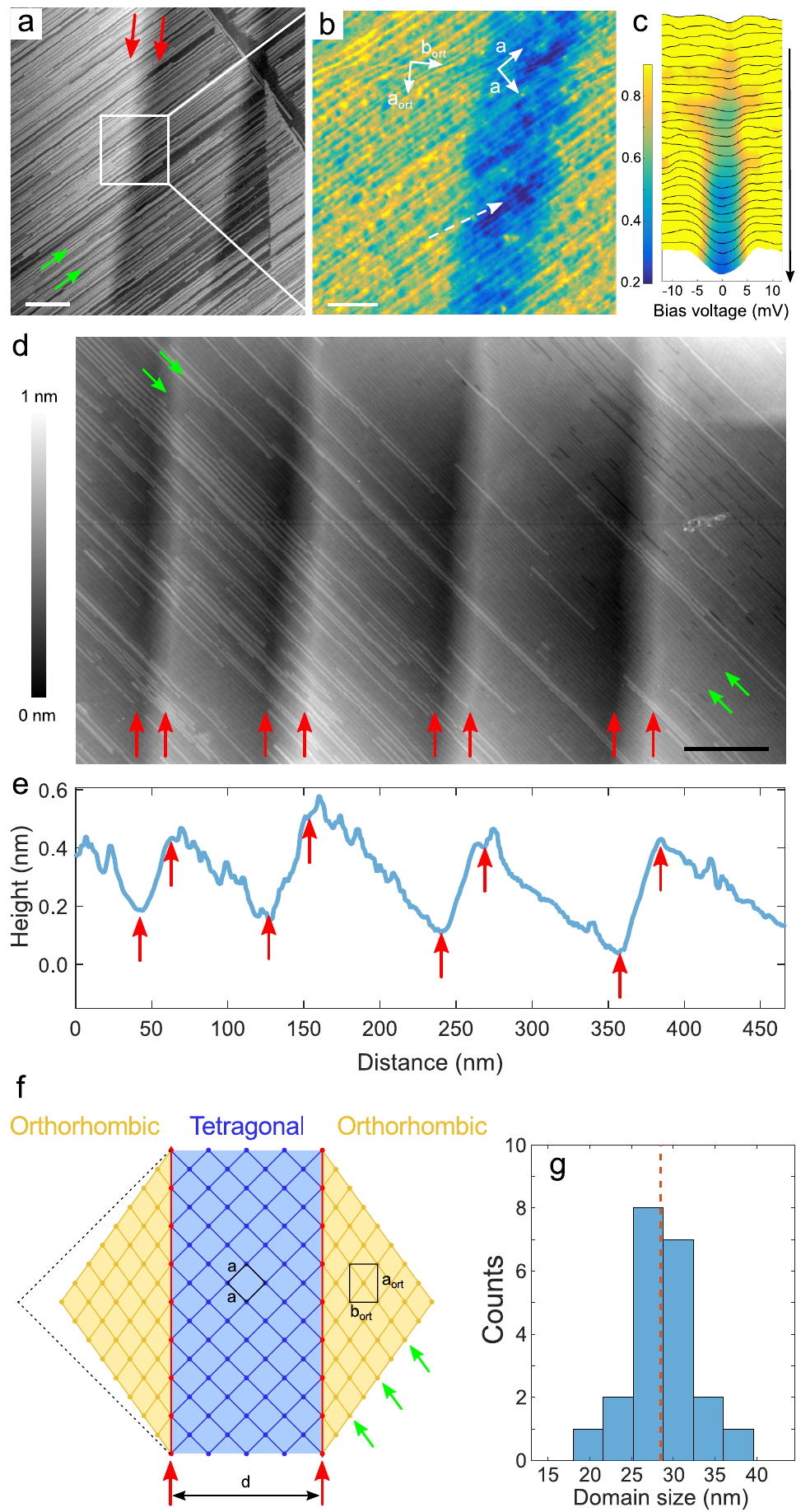}
		\end{center}
		\vskip -.4 cm
		\caption{STM on the tetragonal superconducting domains. In (a) we show an STM image with the stripe features characteristic for the domains observed in AFM (Fig.\ref{DiagramaFases}). Color scale from black to white represents $0.8\,$nm and white scale bar is of $40\,$nm. Red arrows mark the tetragonal domains and green arrows the surface reconstruction. In (b) we show a zero bias conductance map in the area marked by the white square in (a). We also represent tetragonal and orthorhombic unit cell directions. Scale bar is of $10\,$nm. The color scale for the conductance map is the same as used in (c), where we plot the full bias dependent tunneling conductance along the white dashed arrow in (b). In (d) we show another field of view. Tetragonal domains are marked again by red arrows and green arrows mark the surface reconstruction. In (e) we show a line scan at the bottom of (d). In (f) we show schematically the domain structure (not to scale). Atomic As lattice is represented by dots and the domain walls by red arrows. In (g) we show a histogram over the size $d$ of the tetragonal domains. Red dashed line marks the center of the histogram.}
		\label{Domains}
	\end{figure}

Using STM measurements we obtain atomic resolution and identify electronic and crystallographic properties of these areas (Fig.\,\ref{Domains}). The STM images show the $2\times 1$ reconstruction covering most of the surface characteristic of the surface of many pnictide compounds and previously found in the same system in the magnetically ordered phase\cite{Hoffman2011,Gao2010,Chuang2010}. The reconstruction is explained in detail in the Appendix and consists of rows of Ca atoms following the underlying As square lattice and the tetragonal unit cell axis (we mark the reconstruction by green arrows in Fig.\,\ref{Domains}(a), (d) and (f)). In addition to the surface reconstruction, we observe striped features all over the image, always oriented at $45^{\circ}$ with the reconstruction (marked by red arrows in Fig.\ref{Domains}).
	
At the stripes (Fig.\ref{Domains}(b)) we observe regions showing superconducting tunneling conductance curves. We find that the size of the superconducting gap extracted from the tunneling conductance matches the expected energy value for the $T_c$ observed using MFM ($\Delta = 1.76\, k_BT_c \approx 2.3\,$mV).

Each time we observe a stripe, we can identify two kinks in the topography. The overall height changes along one stripe are of a fraction of nm (Fig.\ref{Domains}(d) and (e)). As revealed by X-ray scattering, the lattice distortion between tetragonal and orthorhombic phases is such that the diagonal of the in-plane square of the tetragonal unit cell axis is the same as the long axis of the in-plane rectangle of the orthorhombic unit cell axis\cite{Boehmer2017}. With the spatially resolved domain structure we identify here, we can infer the orientation and distribution of domains. We show the result schematically in Fig.\ref{Domains}(f). Domains are oriented with tetragonal and orthorhombic axis rotated by 45$^{\circ}$ to each other. Note in particular that the interface between domains should have no stress within the plane, because atomic positions coincide along the interface.

In Fig.\ref{Domains}(g) we draw a histogram over the lateral sizes obtained in the STM experiments. We see that these are mostly of a few tens of nm, implying elongated superconducting regions whose size is just a few times the coherence length, as we will see below.

The kinks observed in the STM and AFM images can be adscribed to small angular changes of about 1 \% in the $c$-axis orientations of both domains. These present indeed small variations in the $c$-axis lattice parameters that result in the observed kinks at the surface (we explain the details in the Appendix). In turn this shows that, even if there are atomic positions coinciding at the interface (Fig.\,\ref{Domains}(f)), which eliminate in-plane stress in between domains along the interfaces shown in Fig.\,\ref{Domains}(f), the resulting situation does not fully release the stress along the c-axis.

\begin{figure}
		\begin{center}
			\includegraphics[clip=true,width=0.45\textwidth,keepaspectratio]{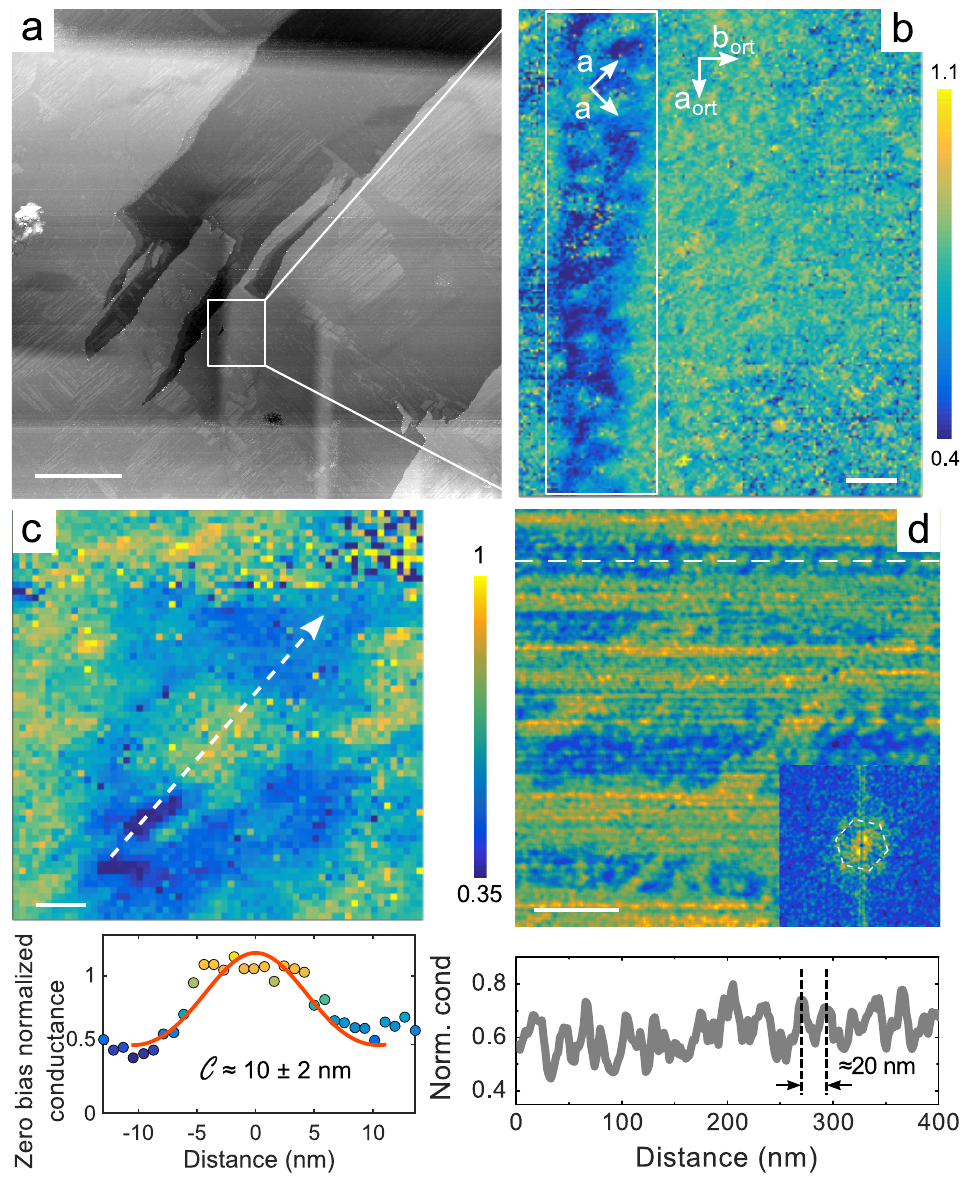}
		\end{center}
		\vskip -.4 cm
		\caption{Vortex lattice in the tetragonal domains. In (a) we show a STM topography image of a large and flat area shown an isolated tetragonal domain roughly at the center. In (b) we zoom into the small white square of (a), holding one single tetragonal stripe. White bar is 300 nm in size. White bar is of 20 nm size. We mark by arrows the orientations of the tetragonal and orthorhombic lattices (see also Fig.\ref{Domains}(f)). The color code of the conductance map is given by the bar on the right. In (c) we show a zoom into a single vortex. White bar is of 5 nm and the color code of the image is given by the bar on the right. The dashed arrow gives the line along which we take the line scan of the bottom panel. The red line in the bottom panel is a fit according to Ref.\cite{Fente2016} (see also text). In (d) we show a zero bias conductance map in an area showing four tetragonal domains that lie close-bye (along the horizontal axis of the image, white bar is of 80 nm), giving a vortex core size $\cal{C}$$\approx 10$ nm. The inset shows a Fourier transform of the image with a white hexagon marking the position of the Abrikosov vortex lattice Bragg peaks expected at this magnetic field (6 T). In the bottom panel we show a line scan along the white dashed line of the top panel. From the line scan we obtain the intervortex distance (20 nm) expected for this magnetic field (6 T).}
		\label{Vortices}
	\end{figure}

When applying a magnetic field, vortices enter these superconducting areas.  In Fig.\,\ref{Vortices} we show results obtained at 6 T. We find the expected intervortex distance for bulk superconductivity at $6$ T indicating we are imaging a bulk vortex lattice. The vortex lattice is quite disordered but there is a clear tendency to have hexagonal like arrangements. Only one or two vortex rows enter in each tetragonal domain. When we zoom into a single vortex (Fig.\ref{Vortices}(c)) we find round vortex cores without identifiable Caroli de Gennes Matricon core states \cite{Caroli1964,Hess1990,Guillamon2008c}. From the profile of the vortex we can estimate a core size $\cal{C}$ of $\approx 10\,$nm using the procedure described in Ref.\citep{Fente2016}. This value is compatible to the measured bulk $H_{c2}\approx 20\,$T if we take into account the expected reduction of the vortex core size with magnetic field\citep{Ran2012,Fente2016}.

Notably, the tunneling conductance maps show lines of high zero bias conductance along the direction of the surface reconstruction. This leads to the yellow lines along the diagonal shown clearly in Fig.\ref{Vortices}(d). Furthermore, the zero bias tunelling conductance never reaches zero in our sample. All this indicates that the surface reconstruction has a strong effect on the tunneling conductance. Likely, there is an associated pair breaking effect related to the $s\pm$ features of superconductivity in this system\cite{Kogan2009,Hirschfeld2011,Chi2016,Chen2016}. As we show in the Appendix, we can identify in some areas the unreconstructed As lattice. These areas are however very small and, although there is a clear pair breaking effect observed on top of lines of reconstructed Ca atoms, it is quite difficult to disentangle from features that may result from tunneling into localized electronic levels of the Ca atoms forming the reconstruction.

To characterize the orthorhombic domains, we have searched for fields of view showing the orthorhombic domains over areas that were sufficiently large to make quasi-particle interference (QPI) scattering experiments with enough $k$-space resolution. As we show in the Appendix, the resulting interference pattern and its bias voltage dependence is fully compatible with previous results  in a purely orthorhombic, non-superconducting and unstrained sample\cite{Chuang2010}. Although here the band closes at a somewhat smaller energy than in the latter case, the shape and energy dependence of the scattering features are essentially the same. Furthermore, we can also identify domains between different orthorhombic orientations, at which the orientation of the nematic signal changes, as in Ref.\cite{Chuang2010,Allan2013}.

\section{Discussion}

The system Ca(Fe$_{1-x}$Co$_x$)$_2$As$_2$ brings a clear-cut example for a first order, discontinous phase transition that can be brought down to zero by an external parameter such as pressure or stress\cite{Gati2012}. This provides a first order quantum phase transition that can be analyzed experimentally in depth. Classical (continous) second order phase transitions as a function of temperature are driven by fluctuations, because of the free energy landscape has a flat minimum at the transition. The fluctuations can be characterized by a relaxation frequency $\tau$. Quantum fluctuations have a faster relaxation than temperature, $\hbar\tau>k_B T$. At any finite temperature, the relaxation frequency of fluctuations disappears when approaching the critical temperature. Thus, sufficiently close to the critical temperature, fluctuations are thermal in nature\cite{0953-8984-17-11-031,RevModPhys.79.1015}. For zero temperature phase transitions, as a function of a parameter other than temperature, spatial and time correlations are intimately connected due to quantum mechanics, drastically modifying the scaling properties of the transition\cite{PhysRevB.14.1165}. But quantum effects can also appear at finite temperatures and lead to anomalous behavior close to the transition\cite{PhysRevB.48.7183,Kopp2005,0034-4885-66-12-R01,SachdevBook,Senthil1490,Coleman2005,Continentino2007828,Macek2014302,PhysRevB.48.7183}. Classical (thermally driven) first order (discontinous) phase transitions occur when the system jumps between separated local minima of the free energy landscape. At sufficiently low temperatures, the energy needed to surmount local minima can be far larger than the temperature. The behavior near discontinous phase transitions can be very different than for continuos phase transitions. For instance, the system might jump between different states through quantum fluctuations\cite{0953-8984-17-11-031,RevModPhys.79.1015}. A number of interesting features have been observed near discontinous quantum phase transitions, such as unconventional superconductivity in ferromagnetic heavy fermions or magnetic textures in intermetallic systems\cite{0953-8984-17-11-031}. For example, in the chiral magnet MnSi, the helical magnetic order breaks up into a lattice of skyrmions, whose size is far above interatomic distances and depends on the balance between magnetic and electronic interactions close to the discontinous helical-paramagnetic phase transition\cite{Muhlbauer915}. For superconducting phase transitions, it has been proposed that coexisting normal-superconducting domains might appear close to discontinous phase transition \cite{PhysRevLett.111.057001,PhysRevB.79.060508,Giraldo2015}. Some calculations show that pnictides are located marginally between phase coexistence and separation \cite{PhysRevB.82.014521}. Here we clearly show that the Ca(Fe$_{1-x}$Co$_x$)$_2$As$_2$ has chosen phase separation. This is not a trivial result---in Ba(Fe$_{1-x}$Co$_x$)As$_2$ and in BaFe$_2$(As$_{1-x}$P$_x$)$_2$, by contrast, antiferromagnetism and superconductivity coexist and show continous phase transitions \cite{PhysRevLett.103.087001,PhysRevB.82.014521,Hashimoto1554}. To the best of our knowledge, there has been no report directly visualizing phase separation at a discontinous phase transition involving superconductivity.

It is important to note that the phase separated state is obtained here by applying strain and not stress nor pressure\cite{Boehmer2017}. Here the sample length is imposed by the substrate. It is useful to repeat the simplified picture provided by authors of Ref.\cite{Boehmer2017}, which bears some analogy to the picture we have of the intermediate state of superconductors\cite{Brandt2011,Prozorov2008}. If one applies a magnetic field to a superconducting sample with a shape providing large demagnetization in the Meissner phase, the edges of the sample transit to the normal phase when the critical field is reached at that location. This redistributes the magnetic field lines so that the field does not increase above the critical field at the interfaces between normal and superconducting areas, giving stable coexisting superconducting and normal regions. In Ca(Fe$_{1-x}$Co$_x$)$_2$As$_2$, when the sample is deformed in-plane by the substrate's differing thermal expansion, the $c$ axis expands as compared to free standing samples. This makes the orthorhombic phase thermodynamically more favorable, because the in-plane length is smaller than the tetragonal phase. At some value of the deformation, orthorhombic domains appear. This reduces the deformation, so that the remaining tetragonal domains stay stable. 

Quite likely, the size of the domains can be modified by applying uniaxial stress to the substrate, either perpendicular or parallel to the stripes. Or simply by changing the substrate. For instance, the thermal expansion of glass is of -0.1\% which should result in a differential thermal expansion of 0.6\% between sample and substrate and eventually lead to modified length scales in the domain size and distribution. Thus, strain might be used as a control parameter to produce novel kinds of superconducting systems, such as intrinsic Josephson junction arrays or to use the domain structure to improve vortex pinning. At very low magnetic fields we observe sometimes linear diamagnetic structures in the orthorhombic phase that might join elongated tetragonal domains, suggesting that such a coupling between elongated domains can indeed happen in some parts of the sample. A macroscopic hallmark for coupled superconducting regions would be non-linear I-V curves with structures at integer multiples of the superconducting gap\cite{PhysRevB.63.180503}. Contrary to nanofabricated arrays, one can envisage here highly transparent interfaces between superconducting and normal phases.

Furthermore, it would be interesting to analyze theoretically the interface between tetragonal and orthorhomic domains\cite{Hirschfeld2011}. We have not been able to study in detail this interface. We only acquired tunneling conductance curve close to the interface that were strongly influenced by the surface reconstruction, because we did not observe a domain boundary in an area showing an unreconstructed surface. The largest areas showing unreconstructed surfaces we have observed are just 20 nm in lateral size. Eventually, if one can tune the size and position of the domains by applying stress, one might well be able to bring the interface into an unreconstructed area and find situations to study the interface with atomic resolution.

We should note that, at present, twin boundaries between two superconducting domains show either enhanced superconductivity, possibly due to increased spin fluctuations in doped BaFe$_2$As$_{2}$ $s\pm$ superconductor, or practically no influence on superconductivity at all in FeSe, the latter being interpreted as a superconducting order parameter rotating at the boundary between two domains with $d$-wave lobes at 90$^{\circ}$ to each other (and locally breaking time reversal symmetry) \cite{PhysRevB.83.064511,PhysRevX.5.031022}. For the domain boundaries we consider here, one should discuss the proximity effect between superconducting tetragonal and nematic orthorhombic domains. In particular, the central hole pocket transforms at the interface from a circular into a nematic hole band. Nematicity is oriented parallel to the domain boundary and the proximity effect will involve breaking the in-plane symmetry of the superconductor by the two-fold nematicity. In addition, Ca(Fe$_{1-x}$Co$_x$)$_2$As$_2$ is likely a  $s\pm$ superconductor with sign changes between the hole pocket located at the center of the Brillouin zone and the electron pockets at the edges. If the interface produces interband scattering, it may well lead to the formation of localized states. Finally, ortorhombicity is usually accompanied by the onset of antiferromagnetic order. Our domain boundary is located in such a way that the magnetic moments change their sign perpendicular to the interface. Assuming that magnetic order remains till the interface itself, it will lead to equally oriented spins along the interface. Such linear alignment of ferromagnetic spins might also lead to the formation of a localized state.

In summary, we show directly microscopic phase separation associated to the optimal in $T_c$ in pnictides. The likely absence of magnetic order in the tetragonal domains, having in close spatial proximity a magnetically ordered domain, suggests that magnetic and superconducting order are both antagonistic, although they are probably fed by the same fluctuations.

\section{Appendix}	

\subsection{C-axis changes between tetragonal and orthorhombic domains and surface topography}
	
\begin{figure}[htb]
		\begin{center}
			\includegraphics[clip=true,width=0.45\textwidth,keepaspectratio]{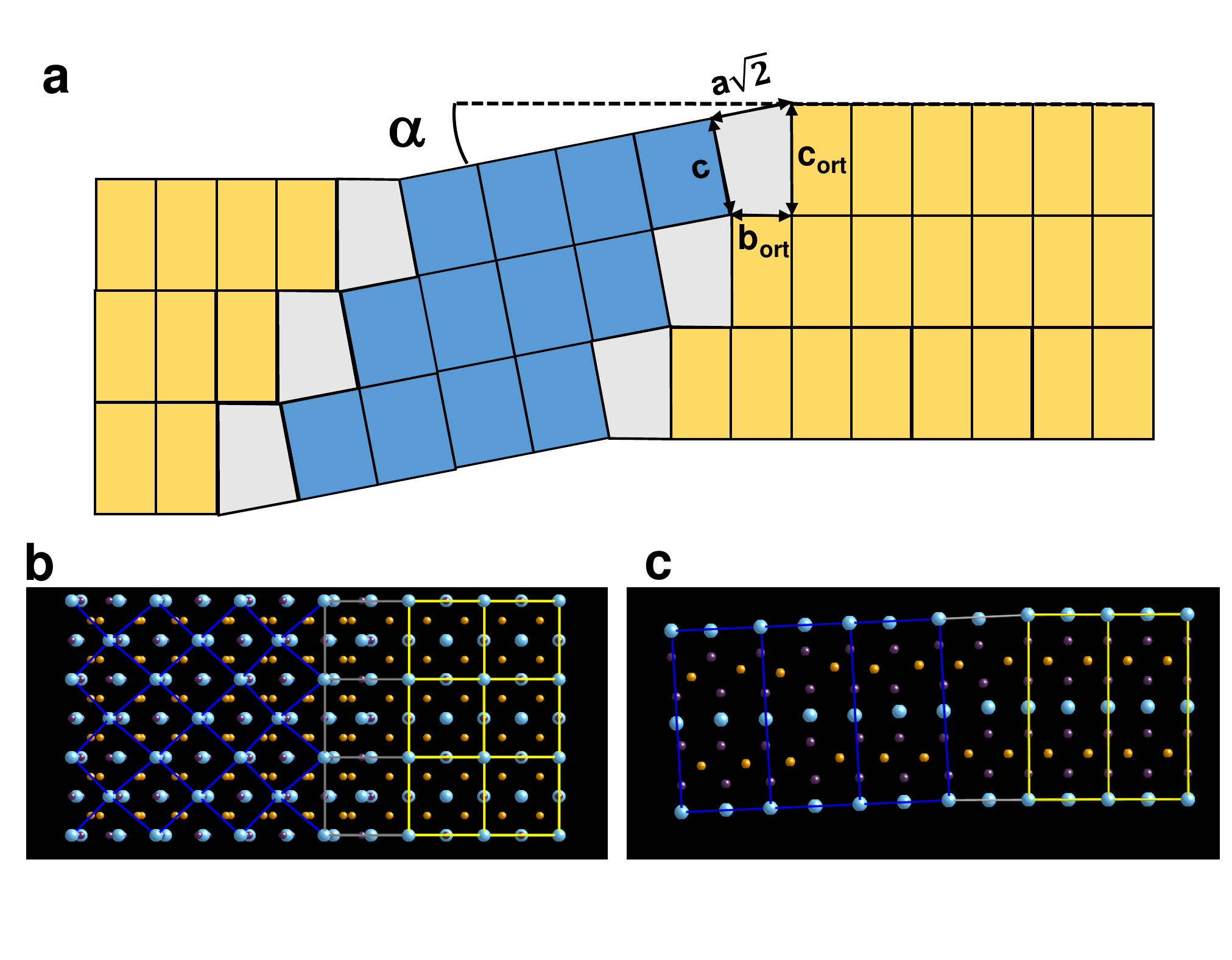}
		\end{center}
		\vskip -.4 cm
		\caption{(a) Schematics (not to scale) of the $c$-axis distortion produced by connecting tetragonal and orthorhombic domains. We schematically show a domain boundary from the side, that is, in the $c$-axis vs in-plane axis plane. Tetragonal domain is marked by blue rectangles and orthorhombic domain by orange rectangles. At the interface, we match the diagonals between both lattices. This provides an angular distortion at the matching line, which leads to a surface that is tilted between both domains. We schematically show the lattice parameters and the tilt angle $\alpha$. The tilt leads to the observed corrugation in AFM and STM measurements. (b) Schematics of the atomic arrangements (not at the right atomic positions neither to scale), viewed from the top, at one interface. The orthorhombic unit cell is marked by yellow rectangles and the tetragonal by blue squares. (c) Same viewed from the side. Ca atoms are in light blue, Fe (or Co) atoms are in orange and As atoms are in purple.}
		\label{cAxis}
	\end{figure}

The corrugation observed in STM and AFM reminds AFM measurements below the tetragonal to orthorhombic transition in BaTiO$_3$ \cite{PhysRevB.86.144416} and STM measurements below the Verwey transition in FeO$_3$\cite{PhysRevB.88.161410}. In both systems, the surface corrugation is associated to a reorientation of the structural domains due to changes in the lattice parameters at a structural transition. We can understand the observed behavior by taking a look on the simplified picture shown in Fig.\,\ref{cAxis}. The condition for matching lattice parameters is met in the plane (Fig.\,\ref{Domains}). But there is also a difference in the c-axis lattice parameters of about 1\% between both phases, so there is in principle no matching along the c-axis \cite{Boehmer2017}. Within a simplified picture, we can however accomodate this distortion along the width of the sample. In Fig.\,\ref{cAxis} we represent the particular case of a highly simplified situation with tetragonal and orthorhombic unit cells sharing one axis (and with the orthorhombic cell volume coinciding with the volume defined by the tetragonal unit cell rotated by 45$^{\circ}$ and expanded in the plane by $\sqrt{2}$). Then, we can match the diagonals of the in-plane vs c-axis rectangles (Fig.\,\ref{cAxis}). The angle formed by the c-axis and in-plane axis of both lattices at each side of the wall is not exactly of 90$^{\circ}$, but differs by $\alpha=90^{\circ}-\arctan\left({\frac{c}{\sqrt{2}a}}\right)-\arctan\left({\frac{b_{ort}}{c_{ort}}}\right)$. Using the lattice constants given by the X-ray data of Ref.\cite{Boehmer2017} ($a_{ort}=5.526$ \AA, $b_{ort}=5.482$ \AA, $\sqrt{2}*a=a_{ort}$ and $c=11.384$ \AA, $c_{ort}=11.575$ \AA), we find $\alpha$ or order of $\approx 0.55^{\circ}$.  Given the crude approximations involved, we believe that the agreement with our experiment, where we consistently find $\alpha \approx 1^{\circ}$ in STM as well as AFM, is quite good.

\subsection{Orthorhombic domains and nematic signal}

\begin{figure}[htb]
		\begin{center}
			\includegraphics[clip=true,width=0.45\textwidth,keepaspectratio]{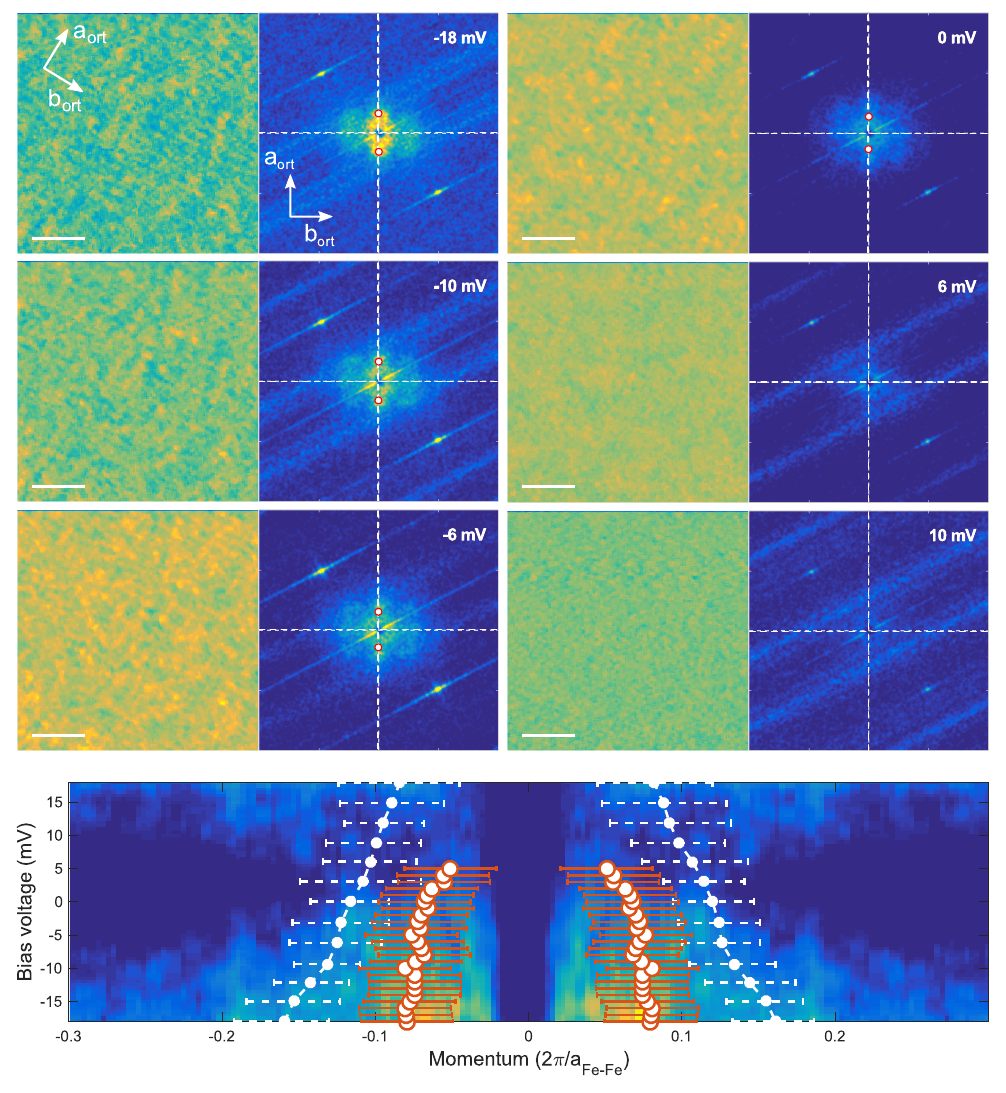}
		\end{center}
		\vskip -.4 cm
		\caption{Tunneling conductance maps at different bias voltages taken in an orthorhombic domain. We show in total six panels consisting of the real space maps (at the left in each panel) and its Fourier transform (at the right in each panel). White scale bars in left panels are of $10\,$nm. The two shiny spots in the right panels are due to the surface reconstruction (see also below). In each panel we show the corresponding bias voltage. In the upper left panel we show the directions of the unit cell. We mark by red dots the position in Fourier space of the hole band for each bias voltage. Our data are without any image treatment. In the bottom panel we show as a color scale the Fourier intensity along the $y$-axis vs the bias voltage. We highlight the dispersion relation of the nematic hole band by red circles. White points give the band dispersion found previously in a fully orthorhombic sample, and are obtained from \citet{Chuang2010}. }
		\label{QPI}
	\end{figure}

In the orthorhombic areas, tunneling conductance curves show no clear signature of superconductivity. In Fig.\ref{QPI} we show the modulation due to quasiparticle interference inside an orthorhombic domain. The modulation provides a central elongated lobe in the Fourier transform where we can identify two main scattering points (along the $y$-axis as shown in Fig.\ref{QPI}) and two reflections at the sides at a distance of $\approx 8 \,a_{Fe-Fe}$. Similar results were previously observed in this material in a purely antiferromagnetic/orthorhombic sample, and are due to a nematic hole band at the $\Gamma$ point \citep{Chuang2010,Allan2013}.

\begin{figure}[htb]
		\begin{center}
			\includegraphics[clip=true,width=0.45\textwidth,keepaspectratio]{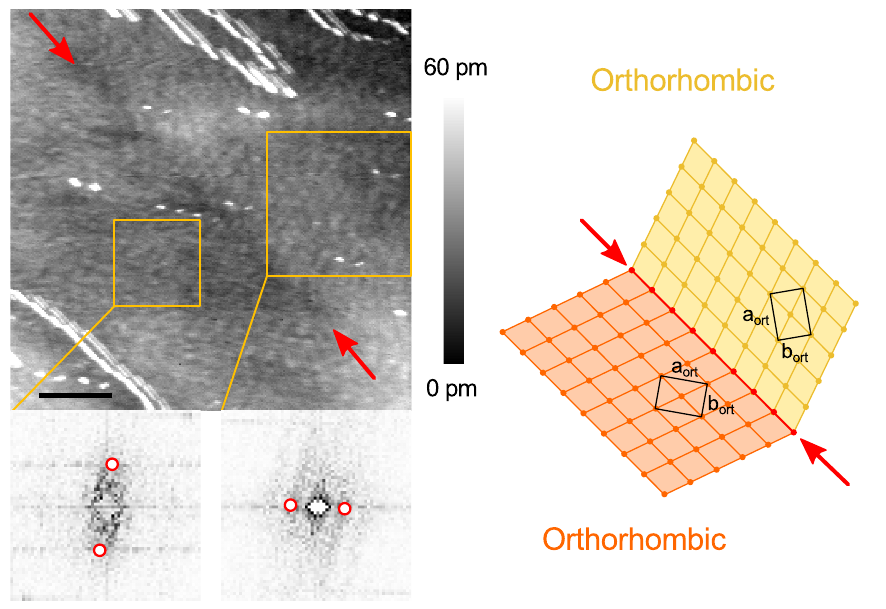}
		\end{center}
		\vskip -.4 cm
		\caption{STM topography over an orthorhombic twin boundary (color scale shown by the bar on the right), marked by two red arrows. 2D-FFT of each part of the image (marked by the two orange squares) show the different directions of the nematic signal (marked by red points as in Fig.\ref{QPI}). Right panel shows schematically the boundary between the two domains and its relative orientation with respect to the crystal lattice (marked as black rectangles). Black scale bar in upper left panel is of $30\,nm$.}
		\label{NematicDomains}
	\end{figure}

The nematic electronic structure has two domains, corresponding to two orthorhombic domains as shown in Fig.\ref{NematicDomains}, again similarly as in \citep{Chuang2010,Allan2013}.

\subsection{MFM images at low magnetic fields}

	\begin{figure}[htb]
		\begin{center}
			\includegraphics[clip=true,width=0.45\textwidth,keepaspectratio]{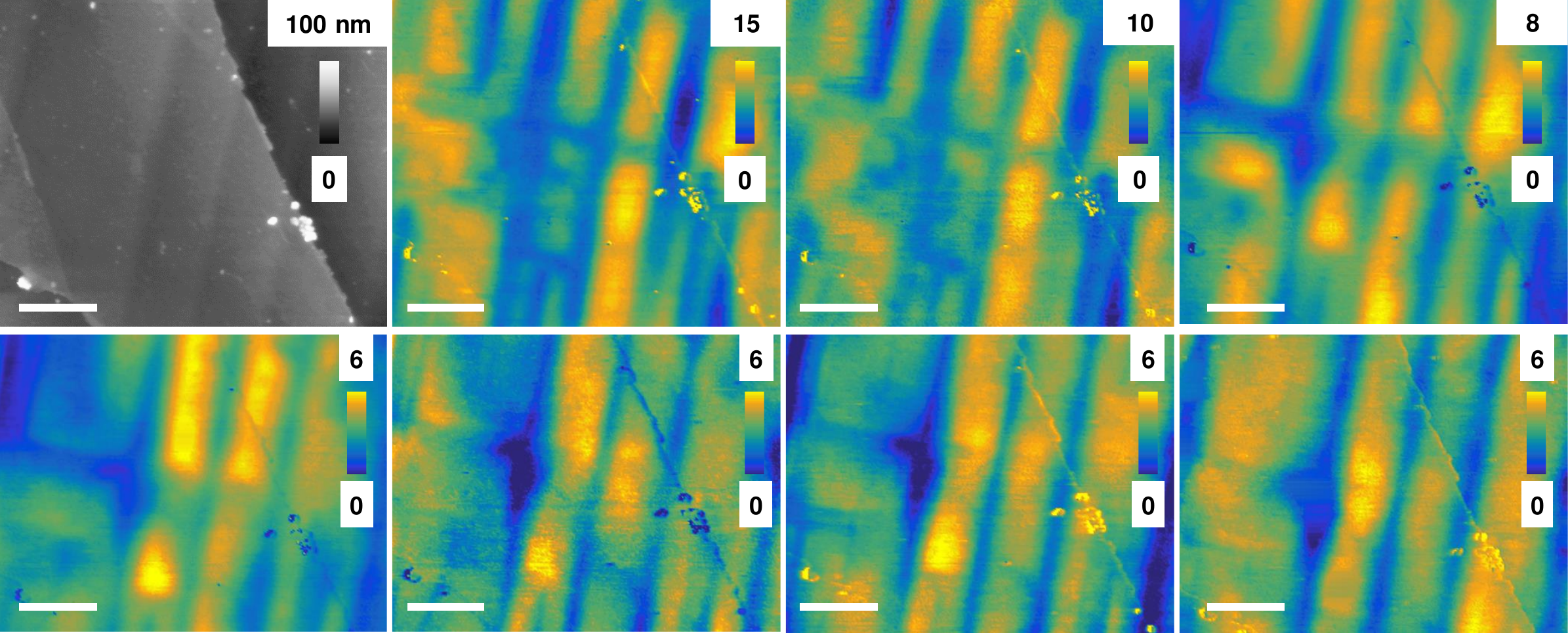}
		\end{center}
		\vskip -.4 cm
		\caption{Evolution of the diamagnetic signal at very low magnetic fields. In the upper left panel we show the AFM topography of the subsequent images. Corrugation is given in nm by the color bar. White scale bar is of 1$\mu$m. In the rest of the panels we show the evolution of the MFM images when increasing the magnetic field at 4 K (images taken at 2.5, 7, 30, 43, 70, 116 and 136 mT). The scale obtained from the MFM is given, in arbitrary units, by the color scales at each panel. The area remains the same over the whole field sweep.}
		\label{MagField}
	\end{figure}

Remarkably, at some locations, the MFM images show linear structures, coinciding with the linear structures highlighting tetragonal domains observed in the topography, interspersed with features oriented perpendicular to the linear tetragonal domains (Fig.\,\ref{MagField}). Note that the resolution of the MFM is not enough to obtain isolated vortices within such a non-uniform magnetic signal. When increasing the magnetic field we observe that the overall difference between large and small magnetization decreases and that the perpendicular domains become normal (Fig.\,\ref{MagField}(b)-(h)). In principle, this could hint to tetragonal domains that join perpendicular to each other, close to an orthorhomic domain boundary. However, that would also result in surface corrugation, which we do not observe (Fig.\,\ref{MagField}(a)). Thus, it seems that fluctuations or disorder might induce superconducting correlations within some areas of the orthorhombic phase.

\subsection{Superconductivity and surface reconstruction}

\begin{figure}[htb]
		\begin{center}
			\includegraphics[clip=true,width=0.45\textwidth,keepaspectratio]{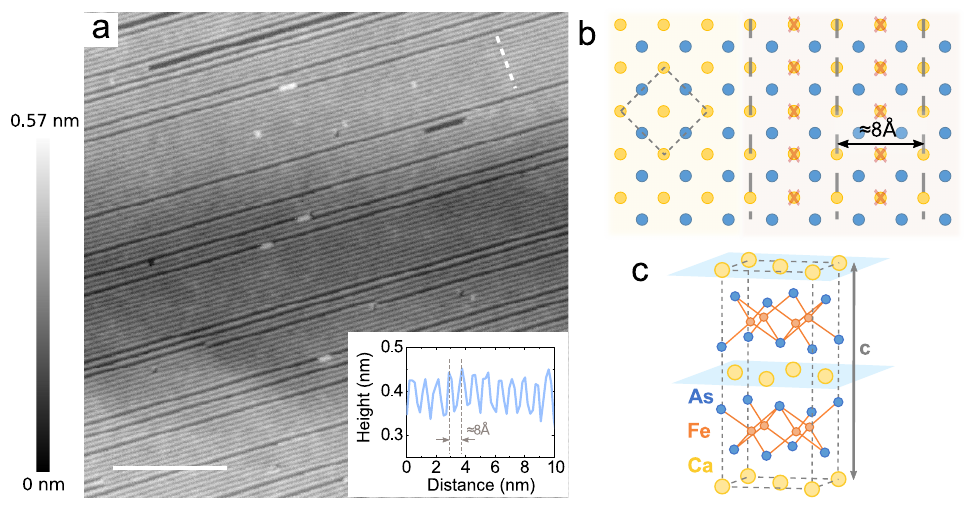}
		\end{center}
		\vskip -.4 cm
		\caption{In (a) we show a topographic image where we can clearly identify the $2\times 1$ reconstruction in detail. White scale bar is of $20\,nm$. Inset contains the profile over the white dashed line (top right part of the image) showing the spacing between the rows forming the reconstruction. In (b) we schematically show the top view of the crystalline structure. The right part of the image shows how the reconstruction is formed along the axis of the Ca-Ca lattice by removing every other row of Ca atoms in the surface. In (c) we show the crystal structure of CaFe$_2$As$_2$. Cleaving planes are highlighted in light blue.}
		\label{Reconstruction}
	\end{figure}

We observe the surface of Ca(Fe$_{1-x}$Co$_x$)$_2$As$_2$ completely covered by a $2\times 1$ surface reconstruction (Fig.\ref{Reconstruction}(a)). This reconstruction is formed when cleaving at low temperatures. Cleaving occurs in the Ca plane (highlighted in light blue in Fig.\ref{Reconstruction}(c)) in a way that half of the layer remains in each one of two remaining parts of the sample. The Ca atoms arrange into a $2\times 1$ reconstruction forming 1D rods separated by $\approx 0.8\,nm$ at an angle of $45^{\circ}$ with the crystallographic axis\cite{Hoffman2011,Gao2010} (see the sketch in Fig.\ref{Reconstruction}(b)).

	The surface reconstruction can be oriented along the two directions the Ca sublattice of the crystalline structure and covers most of the surface of the sample. Thanks to our in-situ system allowing to change in-situ the scanning window, we were able to find few small areas, of the order of a few tens of nm where the reconstruction is absent. These areas usually appear close to borders of the two domains of the reconstruction. At these places, we observe the sublying As atomic lattice. Probably, this situation is metastable and results from the energy cost in establishing the two equivalent surface domains of the reconstruction.

	\begin{figure}[htb]
		\begin{center}
			\includegraphics[clip=true,width=0.45\textwidth,keepaspectratio]{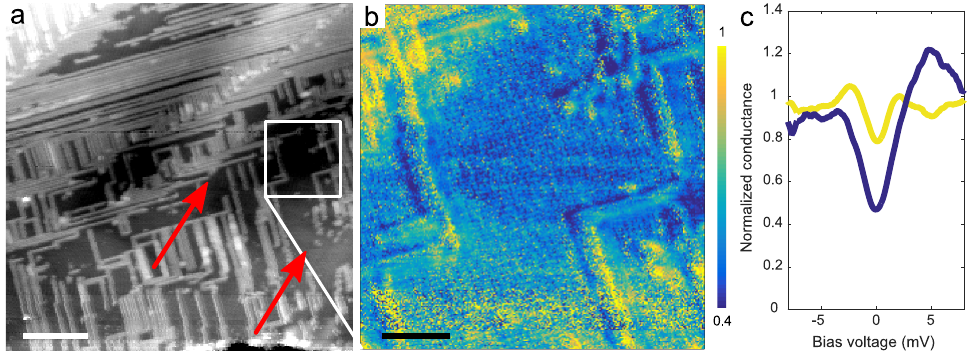}
		\end{center}
		\vskip -.4 cm
		\caption{In (a) we show  a topographic image of a series of tetragonal domains in an area where the As atomic lattice is exposed. White scale bar is of $100\,nm$. In (b) we zoom into one of the tetragonal domains free of reconstruction (marked by a white square in (a)) and show in detail a zero bias conductance map over this area. Black scale bar is of $20\,nm$. The color scale of the conductance map is given by the bar at the right. In (c) we show the normalized tunneling conductance curves taken over a line of Ca atoms (yellow) and on the As lattice, away from the reconstruction (dark blue).}
		\label{PairBreaking}
	\end{figure}
	
We have made tunneling conductance maps in areas where the reconstruction is absent in a tetragonal domain (Fig.\ref{PairBreaking}). We observe that the zero bias conductance has a higher value over lines of Ca atoms than on the As lattice. This suggests that the atomic Ca rows forming the reconstruction break pairs, probably strongly influencing the in-gap conductance all over the tetragonal domains. Curves are however quite featured at the same energy scale as the superconducting gap, and it is difficult to disentangle possible features in the curves from electronic states at the Ca atomic rows and from the underlying superconductivity. Remarkably, the zero bias conductance shows an atomic size modulation (Fig.\ref{PairBreaking}(b)). Such atomic size modulations can be explained by atomically varying tunneling matrix elements that result in slight changes in the contribution to the tunneling conductance of different part of the Fermi surface. These modulations are characteristic of superconductors with different gap values over the Fermi surface\cite{Guillamon2008c}.

\subsection{Gap size from tunneling conductance}

\begin{figure}[htb]
		\begin{center}
			\includegraphics[clip=true,width=0.45\textwidth,keepaspectratio]{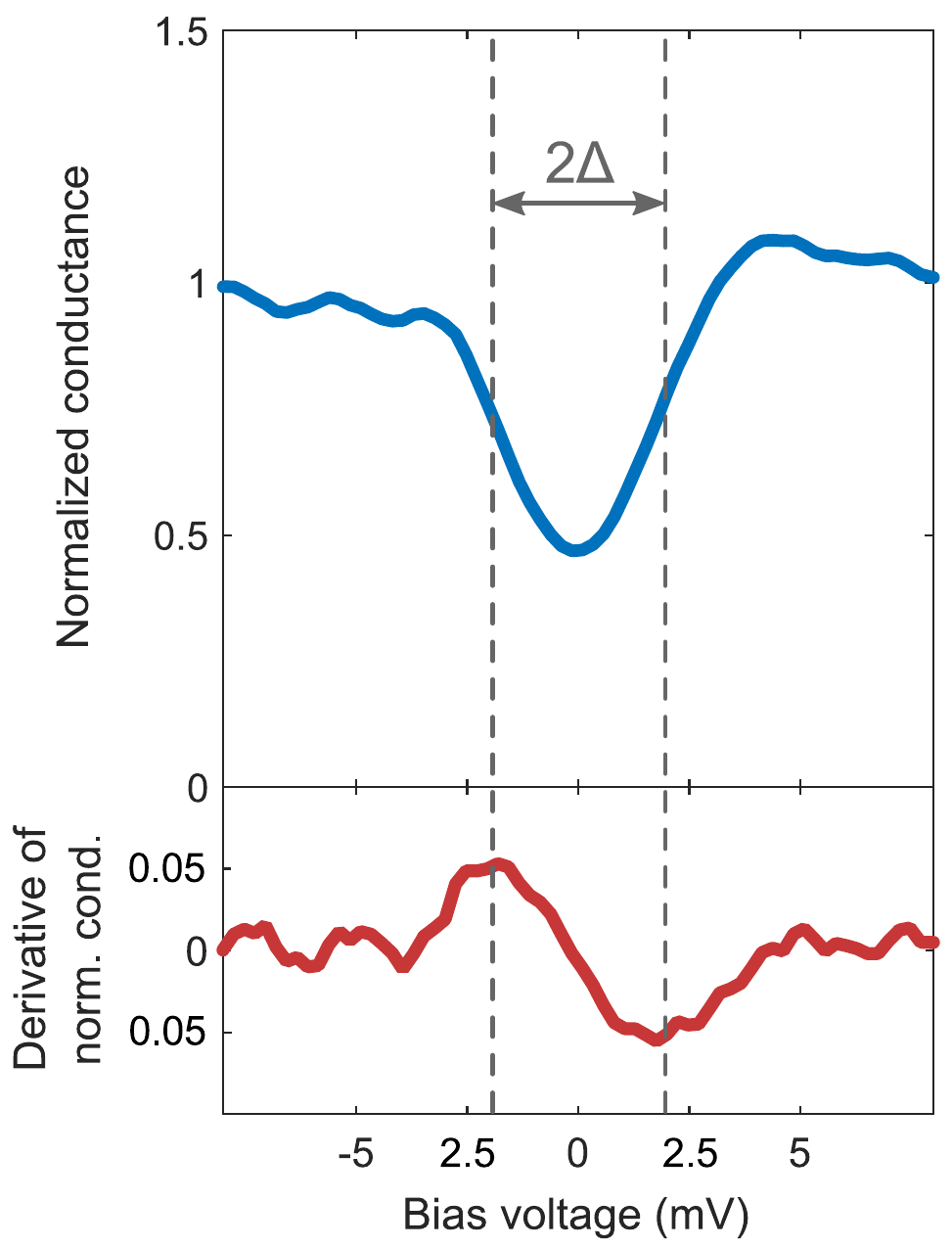}
		\end{center}
		\vskip -.4 cm
		\caption{In (a) we show (blue line) the normalized tunneling conductance obtained typically in tetragonal domains. In (b) we show (red line) its derivative. We mark the BCS value of the superconducting gap ($2\Delta$), obtained using the $T_c$ of our sample extracted from macroscopic magnetization and local scale MFM measurements by dashed lines.}
		\label{Gap}
	\end{figure}

Given that the tunneling conductance is so far from the simple $s$-wave BCS tunneling density of states, we simply obtain the value of the superconducting gap by looking at the maximum of the slope of the tunneling conductance. As we show in Fig.\ref{Gap}, the value thus obtained is the same as the BCS expectation if we use $T_c$ from magnetization and MFM measurements.

\section{Acknowledgments}
\begin{acknowledgments}
Work done in Madrid was supported by the Spanish Ministry of Economy and Competitiveness (FIS2014-54498-R, MDM-2014-0377, MAT2014-52405-C2-2-R, RYC-2014-16626 and RYC-2014-15093), by the Comunidad de Madrid through program Nanofrontmag-CM (S2013/MIT-2850), by the European Research Council PNICTEYES grant agreement 679080, by FP7-PEOPLE-2013-CIG 618321, by EU Flagship Graphene Core1 under Grant Agreement n° 696656 and by Axa Research Fund. SEGAINVEX-UAM is also acknowledged. Work done in Ames Lab was supported by the U.S. Department of Energy, Office of Basic Energy Science, Division of Materials Sciences and Engineering. Ames Laboratory is operated for the U.S. Department of Energy by Iowa State University under Contract No. DE-AC02-07CH11358.
\end{acknowledgments}

\end{document}